# Quantifying Temperature-dependent Substrate Loss in GaN-on-Si RF Technology

Hareesh Chandrasekar, *Member, IEEE*, Michael J. Uren, *Member, IEEE*, Michael A. Casbon, *Member, IEEE*, Hassan Hirshy, *Member, IEEE*, Abdalla Eblabla, Khaled Elgaid, *Senior Member, IEEE*, James W. Pomeroy, Paul J. Tasker, *Fellow, IEEE* and Martin Kuball, *Senior Member, IEEE*

*Abstract*—Intrinsic limits to temperature-dependent substrate loss for GaN-on-Si technology, due to the change in resistivity of the substrate with temperature, are evaluated using an experimentally validated device simulation framework. Effect of room temperature substrate resistivity on temperature-dependent CPW line loss at various operating frequency bands are then presented. CPW lines for GaN-on-high resistivity Si are shown to have a pronounced temperature-dependence for temperatures above 150°C and have lower substrate losses for frequencies above the X-band. On the other hand, GaN-on-low resistivity Si is shown to be more temperature-insensitive and have lower substrate losses than even HR-Si for lower operating frequencies. The effect of various CPW geometries on substrate loss is also presented to generalize the discussion. These results are expected to act as a benchmark for temperature dependent substrate loss in GaN-on-Si RF technology.

## I. INTRODUCTION

GaN-ON-SI MMIC high-power amplifiers are technologically promising for cost-sensitive applications such as wireless base stations and CATVs amongst others. The widespread adoption of this technology however depends upon the availability of high-performance, low-loss active and passive components, which remain subjects of active investigation.[1-4] For instance, even though GaN-on-Si RF components are almost exclusively fabricated on high resistivity Si (HR-Si, ρ>5 kΩ.cm) to minimize substrate conduction losses, issues such as the formation of a parasitic conduction channel along the III-nitride – Si interface are known to increase RF losses and have been studied and modeled.[5-8]

Another critical factor for GaN-on-Si technology is expected to be the impact of high temperatures on the HR-Si substrate resulting in thermal generation of carriers, thus lowering its resistivity and increasing substrate loss.[9] Such higher temperatures could be either due to the application-specific rated operating temperatures for MMICs (nominal maximum case operating temperatures of 85°C -150°C),[10] and/or heat spreading resulting from localized self-heating in the AlGaN/GaN channel of active devices. Since channel temperatures for GaN HEMTs have been measured to be high ($T_{rise}$≤200°C),[11] heat flow from the channel into the epitaxial stack and Si can raise substrate temperatures locally under the device to greater than 150°C at the Si surface (as shown in Appendix A, Fig. 9). These temperatures are not insignificant and their contribution to loss at GHz frequencies has not been well investigated in literature despite their relative importance.[12] Furthermore, the temperature dependence of silicon resistivity is not monotonic and depends on the starting substrate resistivity (Fig. 10 Appendix B).

Transmission line loss is a convenient indicator to quantify such inherent loss mechanisms, and state-of-the-art conventional co-planar waveguide (CPW) transmission line losses down to 0.8 dB/mm at 110 GHz have been demonstrated for GaN on HR-Si at room temperature.[4, 13]. It must be mentioned that the effect of temperature on transmission line loss is well recognized in bulk silicon RF technology and potential alternatives such as silicon-on-insulator, porous and trap-rich Si substrates have been shown to exhibit a higher tolerance to temperature effects.[13-17] However, the integration of GaN with these alternative substrates is hardly straightforward in view of the challenges in epitaxy and processing, to say nothing of potential thermal management considerations.

In this paper, to the exclusion of other factors, we quantify the intrinsic temperature-dependent substrate loss in CPW lines for GaN-on-Si technology using a device simulation model benchmarked to measurements. A comparison between frequency- and temperature-dependent loss for various substrate resistivities is presented for an informed application-specific choice of substrate. These results are expected to serve as a benchmark for temperature-dependent RF substrate

This work was funded by the UK EPSRC under grants EP/N031563/1, EP/N014820/2 and partly by the Engineering Research Network Wales (project NRNC19).

H. Chandrasekar, M.J. Uren, J.W. Pomeroy and M. Kuball are with the Centre for Device Thermography and Reliability, H. H. Wills Physics Laboratory, University of Bristol, Bristol BS8 1TL, UK (e-mail: h.chandrasekar@bristolac.uk).

M.A. Casbon and P.J. Tasker are with the Centre for High Frequency Engineering, School of Engineering, Cardiff University, Cardiff CF24 3AA, UK.

H. Hirshy is with the Compound Semiconductor Applications Catapult, Regus House, Falcon Drive, Cardiff Bay, Cardiff CF10 4RU, UK.

A. Eblabla and K. Elgaid are with the Department of Electrical and Electronic Engineering, School of Engineering, Cardiff University, Cardiff CF24 3AA, UK.

losses for various substrate resistivities in GaN-on-Si technology.

## II. SIMULATION METHODOLOGY AND EXPERIMENTAL DETAILS

The Silvaco ATLAS device simulator was used to perform two-port small-signal a.c. simulations on a generic 50 Ω GaN-on-Si CPW structure (see Fig. 1(a)), consisting of a Si substrate followed by an AlN nucleation layer, linearly graded AlGaN strain-relief layer and highly-resistive GaN buffer, with smooth alignment of bands at AlN/graded-AlGaN and graded-AlGaN/GaN hetero-interfaces and flat band condition at Si/AlN interface.[18, 19] No bulk polarization, doping or trap levels in the III-nitride layers were considered for ease of analysis and so the Fermi energy lies in the mid-gap for these layers.[20] Conductor lines of 0.4 µm thickness were taken to be deposited on a $SiN_x$ passivation layer on top of the GaN film, analogous to the mesa isolation floor for fabricating integrated passive components. No parasitic channel at the Si/AlN interface was considered to focus solely on evaluating temperature-dependent loss in Si due to thermally induced free carriers. The simulation outputs the small-signal capacitances and conductances at each terminal and the associated two-port S-parameters. To account for resistive losses in the substrate, the loss tangent (tan δ = Real($Y_{11}$)/Imag($Y_{11}$) from the Y-matrix) in the structure was extracted (see lumped equivalent circuit shown in Fig. 1(a)). The simulation accounts for the temperature-dependent carrier density, band-gap and mobility effects, all of which affect substrate resistivity. Substrate loss was calculated in a transmission line by treating the substrate as a lossy dielectric. For a given loss tangent, extracted from the small signal network parameters, dielectric loss ($\alpha_d$ in dB/mm) for a CPW line with a center conductor width of $W_C$, gap spacing of S and substrate thickness $t_{sub}$ is given by,

$$\alpha_d = 8.686 \frac{q \varepsilon_r \tan \delta}{\varepsilon_{eff} \lambda_g}, \quad (1)$$

where $\varepsilon_r$ is the relative permittivity of the substrate, $\lambda_g$ is the guide wavelength given by $c/f\sqrt{\varepsilon_r}$, q is the dielectric filling factor calculated according to standard CPW line expressions.[21]

In order to benchmark this device simulation framework, CPW lines were fabricated on $SiN_x$ films deposited on Si substrates of different resistivities. $SiN_x$-on-Si samples were used as they provide a cleaner interface to the substrate without the parasitic conduction channel frequently present in GaN-on-Si samples which would impact the substrate effect being studied here.[7, 8] This allows for the validation of the simulations employed here which can then be extended to GaN-on-Si CPW lines. Two Si substrates (high-resistivity of >5 kΩ.cm and low-resistivity of 10-100 Ω.cm) were used for this study. Electron beam lithography was used to define the transmission line fabrication on these samples. The 160 nm $SiN_x$ passivation layer was deposited using plasma enhanced chemical vapor deposition (PECVD). 50 Ω CPW transmission lines (0.5 mm and 1 mm lengths) were then fabricated on top of this $SiN_x$ layer by depositing Ti/Au (20/450 nm) with the line dimensions as shown in Figs. 1(b) and (c). The small signal S-parameters of the CPW lines were measured from 25°C to 200°C using a Rhode and Schwarz ZVA-67 vector network analyzer mounted on a Cascade Summit 12000M wafer probe station, fitted with Cascade GSG Z-probes.

### III. RESULTS AND DISCUSSION

*A. Temperature Dependent Line Loss for $SiN_x$-on-Si CPWs – Simulation and Experiment*

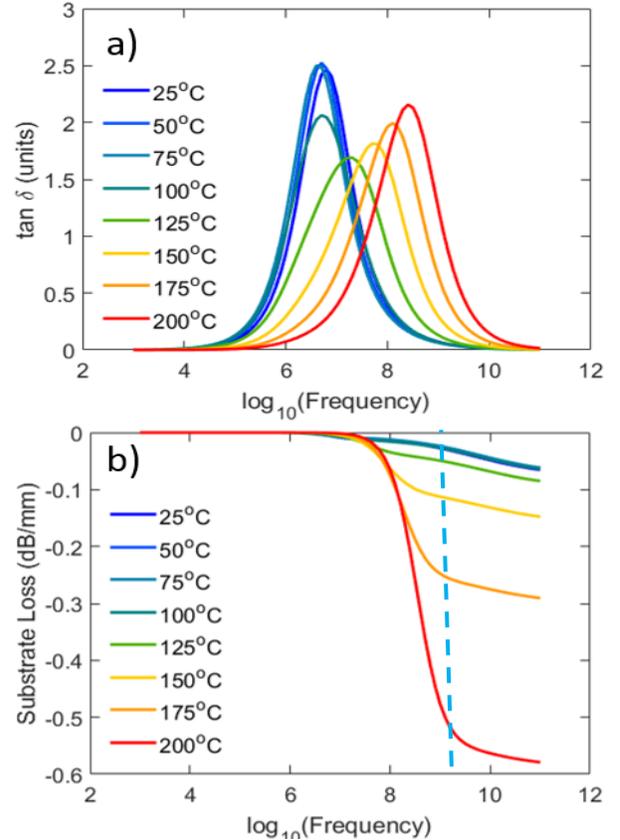

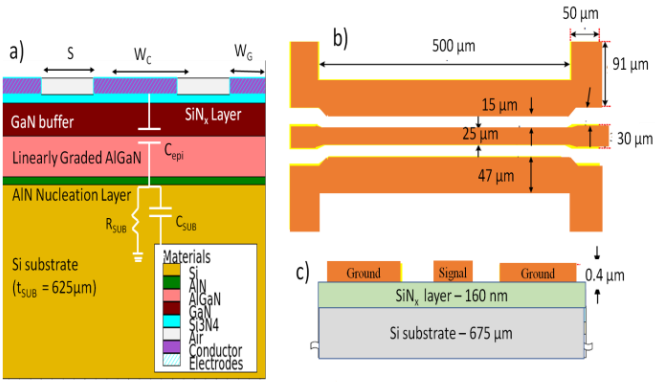

Fig. 1. (a) Cross-section of the simulated CPW structure on GaN-on-Si with representative dimensions and equivalent lumped R-C network at the central conductor. (b) Top view and (c) cross-sectional view of 50Ω CPW lines fabricated on $SiN_x$ films on various Si resistivity substrates to benchmark the device simulations.

Fig. 2. (a) Simulated loss tangent and (b) substrate loss (dB/mm) of CPW lines for $SiN_x$-on-HR-Si (5 kΩ.cm) vs frequency for temperatures between 25-200°C showing pronounced temperature dependence of substrate loss. Dashed vertical line indicates 20 GHz frequency used for experimental comparison.

Figs. 2 and 3 show the simulated loss tangent and substrate loss for CPW lines on 160 nm thick $SiN_x$ layer on HR (5

kΩ.cm) and LR-Si (25 Ω.cm) substrates. The loss tangent peak indicates the frequency at which the capacitive contribution to substrate admittance equal its parallel conductance component (see equivalent circuit in Fig. 1(a)). Since the substrate resistivity and therefore its conductance depends on the temperature, the loss tangent peak shifts in frequency as a function of temperature. The substrate loss increases as the peak shifts right, as it is multiplied by the frequency (inversely proportional to $\lambda_g$ in (1)). We see that substrate loss on HR-Si (Fig. 2(b)) rises for temperatures beyond 80°C due to thermal generation of carriers and increases dramatically beyond 100°C to 0.13 dB/mm at 150°C and 0.56 dB/mm at 200°C at a frequency of 5 GHz for these CPW dimensions (with mild increases beyond), exhibiting a pronounced temperature dependence. On the contrary, simulated loss tangent peaks for CPW lines on LR-Si do not show the same pronounced shifts with temperature and shift to lower frequencies at higher temperatures as shown in Fig. 3(a). This is consistent with the fact that the resistivity of the silicon increases with increasing temperature for LR substrates (due to reduction in carrier mobilities). Notice that this left-shift leads to two frequency regimes of loss tangent (and hence substrate loss) behavior with temperature – in region 1, the loss tangent and substrate loss increase monotonically with temperature for a given frequency <1 GHz, and for frequencies >1 GHz (region 2), a cross-over occurs in the temperature dependence, and both tan δ and loss decrease with increasing temperature.

200°C. The loss tangent peaks are relatively temperature insensitive and left-shift lowering loss with increasing temperature for frequencies in the GHz range – region 2. A cross-over in temperature dependent loss is seen at lower frequencies where the substrate loss increases with temperature - region 1. Arrows indicate trends corresponding to increasing temperature. Dashed vertical line indicates 20 GHz frequency used for experimental comparison.

In order to verify these predicted trends and CPW loss values, small-signal insertion losses were measured on the CPW lines. Fig. 4 shows the measured insertion loss which consists of substrate, conductor and radiation loss contributions, compared to the simulated substrate loss for the two substrates at 20 GHz. Firstly, we see that the insertion loss for the HR-Si increases with temperature, while CPW line loss on LR-Si decreases with increasing temperature – a trend which is entirely captured by the simulations. These observations are in line with the change in substrate resistivities with temperature for high-res and low-res silicon (Appendix B). Secondly, the simulated losses are lower than measured insertion losses as they only represent the substrate contributions to the total line loss at these temperatures. Thirdly, at 20GHz the line loss on HR-Si is lower for the entire range of temperatures compared to LR-Si for these starting substrate resistivities due to the actual position of the loss tangent peak with temperature (Figs. 2, 3).

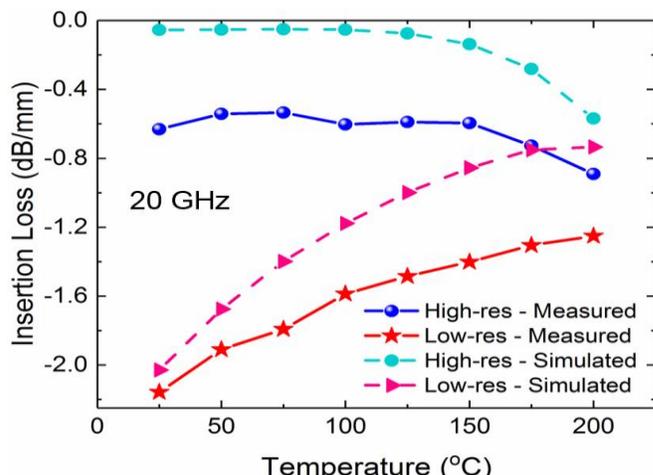

Fig. 4. Measured insertion loss (dB/mm) and simulated substrate loss (dB/mm) at 20 GHz for 160 nm $SiN_x$ on HR- and LR-Si substrates. Both simulations and experiment capture the trend of loss increasing with temperature for high-res silicon and vice-versa for low-res silicon, for reasons discussed in the text.

The insertion loss was then measured at lower frequencies (20 MHz – 6 GHz) on low-res substrates to try and observe the temperature dependent cross-over trend of line loss with frequency discussed above. Fig. 5(a) shows the experimental insertion line loss across this range of frequencies for temperatures from 25-200°C. We observe the same two regimes of line loss increasing with temperature at lower frequencies, followed by a cross-over into a regime where the line loss decreases with temperature, as predicted from simulations, and shown in Fig. 5(b) for comparison. That the experimental and simulated cross-over frequencies also show very good agreement further illustrates that the simulation captures the essential features of this system.

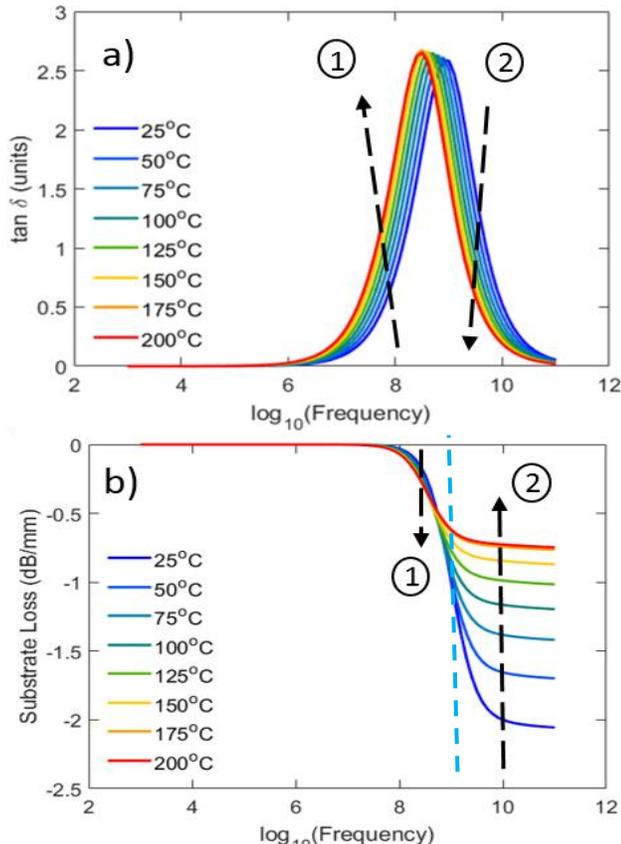

Fig. 3. (a) Simulated loss tangent and (b) substrate loss (dB/mm) of CPW lines for $SiN_x$-on-LR-Si (25 Ω.cm) vs frequency for temperatures between 25-

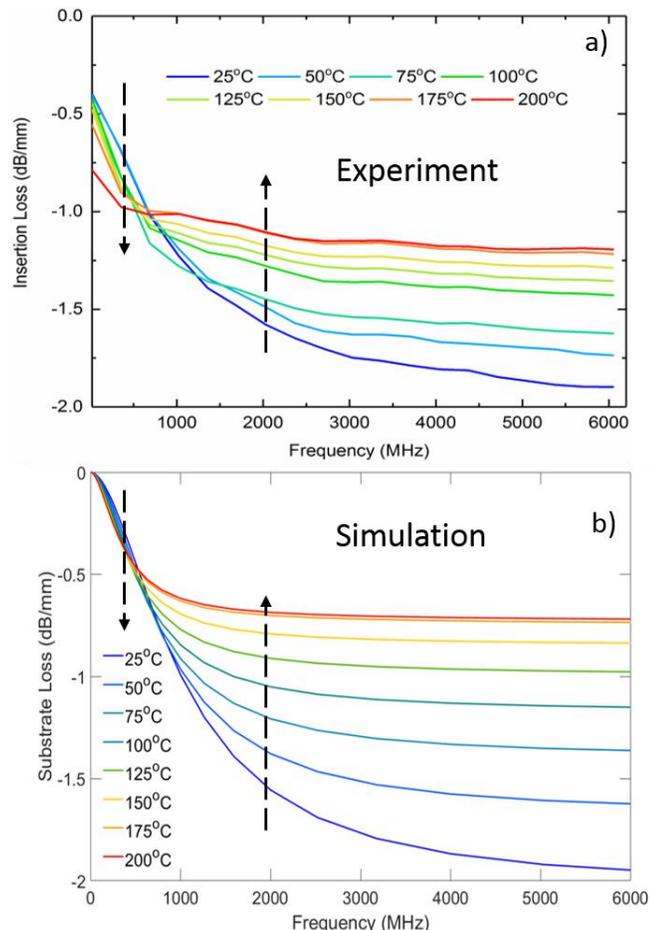

Fig. 5. (a) Measured insertion loss and (b) Simulated substrate loss with frequency for temperatures in the range (25-200°C) for CPW lines on 160 nm $SiN_x$ on low-resistivity silicon showing the cross-over between regions of increasing and decreasing substrate loss with increasing temperatures.

### B. Temperature-dependent Line Loss for GaN-on-Si CPWs - Substrate Resistivity and Operating Frequency Effects

We now simulate temperature-dependent CPW line loss in the prototypical GaN-on-Si epitaxy shown in Fig. 1(a). We have seen that substrate loss depends not only upon the starting substrate resistivity but also on the range of operating frequencies being considered. To better evaluate the effect of temperature and different substrate resistivities on line loss, substrates having high, medium, low and very low resistivities of 10 kΩ.cm, 100 Ω.cm, 1 Ω.cm and 0.01 Ω.cm were simulated having the same generic RF epitaxy on silicon (2.5 µm total thickness, 0.5 µm GaN channel layer with $10^{16}$ cm$^{-3}$ shallow donors and a highly resistive buffer of 2 µm) for comparison. Substrate loss is plotted as a function of temperature for representative microwave frequencies of 2.5 GHz (S-band), 12 GHz (X-band) and 40 GHz ($K_a$-band) in Fig. 6. It is evident that, as far as RF line loss is concerned, medium, low and very-low resistivity Si all offer lower variation with respect to temperature than HR-Si at all three frequency bands. Note that while HR-Si has a lower substrate loss at room temperatures over all frequency ranges (see Figs. 6(a), (b) and (c)) compared to lower resistivity substrates, this is not the case with increasing temperatures. Considering the full range of rated operating temperatures (up to 150°C), very-low resistivity Si (0.01 Ω.cm), counter-intuitively, exhibits much lower substrate loss over the entire range of temperatures at 2.5 GHz compared to HR-Si as seen from Fig. 6(a), in addition to being temperature insensitive.

Furthermore, substrate loss contributions for commonly available low to med-resistivity substrates (1-100 Ω.cm) are much higher than both HR and very-low resistivity substrates. For instance, loss in the 1 Ω.cm substrate is consistently higher than the HR case at 2.5 GHz until a cross-over occurs at temperatures beyond 175°C. Interestingly enough, such a cross-over between GaN-on-HR and LR-Si losses has also been experimentally observed by Takenaka et al. to occur for a 1-port drain pad structure at 150°C, close to these calculated values.[12] On the other hand, HR-Si clearly offers the lowest loss at 40 GHz and above (Fig. 6(c)) for all temperatures when compared to the other substrates while 12 GHz represents an intermediate case between these two frequencies. This discussion is summarized in Fig. 7 where substrate losses vs Si substrate resistivity are plotted at 25°C and 200°C at 2.5 GHz and 40 GHz respectively, clearly indicating that while HR-Si offers much lower loss at higher frequencies (f>12GHz) and has acceptable loss figures even up to 110 GHz, this is not necessarily true for lower frequencies, such as wireless communication in S-band, where highly-doped Si substrates offer an attractive combination of lower absolute substrate loss in addition to temperature-invariance.

We stress once again, that these inferences are based purely on loss in the silicon substrate, and other circuit and device considerations may well determine the final choice of substrate resistivity. For instance, the use of low-res substrates is equivalent to moving the ground plane to the bottom of the epitaxy from that of the Si substrate, and corresponds to higher output capacitance, to say nothing of changing the impedance of the designed 50 Ω transmission line. In order to retain the standard impedance, the CPW dimensions would have to be changed which would in turn alter the substrate loss. Even in terms of temperature-dependence of the total RF loss itself, conductor loss in the CPW lines can be comparable to or worse than substrate losses depending on the geometry as discussed in the next section. Better thermal management strategies to limit the temperature rise in the top surface of the Si substrate to <100°C would also drastically reduce the substrate loss contributions from HR-Si.

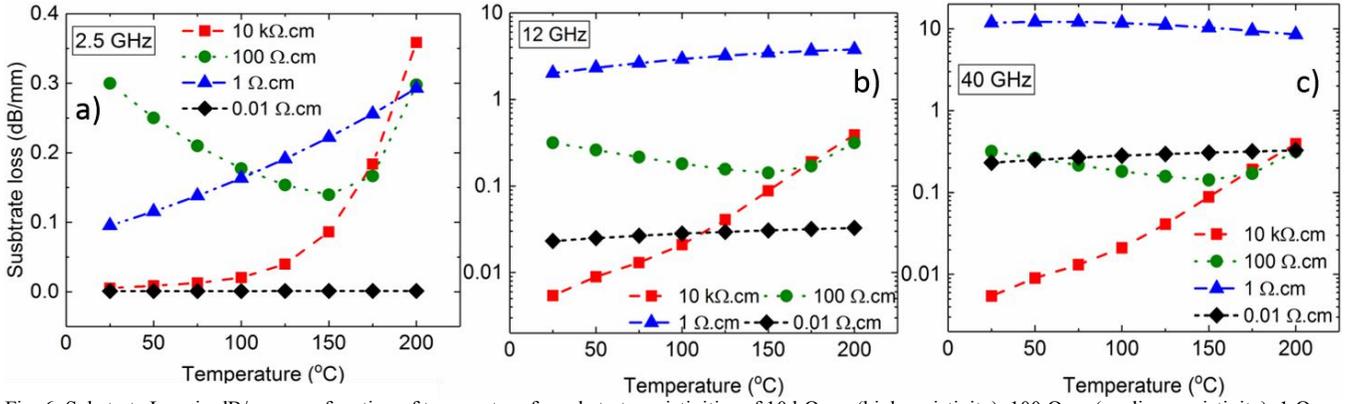

Fig. 6. Substrate Loss in dB/mm as a function of temperature for substrate resistivities of 10 kΩ.cm (high-resistivity), 100 Ω.cm(medium-resistivity), 1 Ω.cm (low-resistivity) and 0.01 Ω.cm(very-low-resistivity) at microwave frequencies of (a) 2.5 GHz, (b) 12 GHz and (c) 40 GHz. Substrate loss is plotted on a log-scale in the latter two plots. High-resistivity Si confers an advantage in terms of lower substrate loss at all temperatures up to 150°C only for f>12GHz, while the very-low resistivity Si has lower loss at reduced frequencies and is more temperature insensitive.

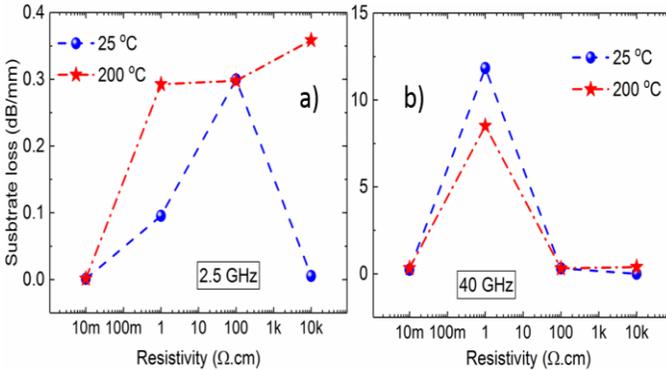

Fig. 7. Substrate Loss (dB/mm) vs starting substrate resistivity at 25°C and 200°C for operating frequencies of (a) 2.5 GHz and (b) 40 GHz. Dotted lines indicate trends between estimated points. Loss is lowest in case of very-low resistivity substrates at lower frequencies, even at 25°C, while HR-Si has lowest losses even at 200°C for 40 GHz operation.

## C. Geometrical Effects on Temperature-Dependent CPW Substrate Loss

We now comment on the effects of CPW geometry and passivation layer thickness on the simulated substrate loss in order to generalize the discussion above. CPW lines of 50 Ω impedance but with 3 different variants of gap spacing (S=7 μm,15 μm and 30 μm) were considered on top of the generic GaN-on-HR Si stack. The substrate loss increases for larger gaps (and hence central conductor widths) due to increased capacitive coupling with the substrate as shown in Fig. 8 at a frequency of 2.5 GHz (note that the variation of loss with frequency for f > 2.5GHz in HR-Si is not significant, as can also be seen from Fig. 6). The trend of substrate loss with temperature remains similar for changes in line geometry, with larger gap spacings showing a higher loss. It is important to note conductor/metal loss in CPW has the converse trend with geometry - it increases with smaller gap spacings. Hence the dominant RF loss might well be due to conductor loss for smaller CPW spacings and not substrate loss.

We also investigated the effect of larger passivation layer thicknesses, to decrease the extent of capacitive coupling between the conductor lines and the substrate. In comparison to the standard nitride thickness of 160 nm used in this work, we simulated the effect of raising the $SiN_X$ thicknesses to 2 μm and 5 μm, for the same CPW line dimensions in Section IV (S=15 μm, $W_C$=25 μm). These results are also summarized in Fig. 8 at 2.5 GHz for temperatures up to 200°C. We see that fabricating a CPW transmission line on top of a 5 μm thick dielectric can cut down the substrate loss at 200°C by 45% as compared to the 160 nm case. In this context the use of alternative strategies such as shielded-elevated coplanar waveguide structures are very promising to reduce substrate resistivity-imposed limitations on line loss by decreasing the capacitive coupling between the lines and the substrate.[22]

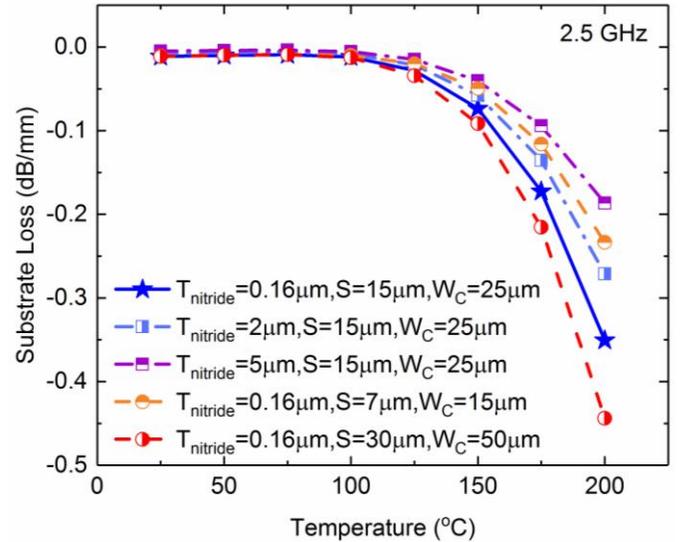

Fig. 8. Comparison of geometry dependence of CPW substrate loss on HR-Si with temperature at 2.5 GHz. Gap spacings of S=7 μm, 15 μm (standard for this study) and 30 μm were considered (squares), along with changes in passivation layer thickness of 0.16 μm (standard for this study), 2 μm and 5 μm (circles). The loss increases with increased coupling to the substrate for higher gap spacings and lower passivation layer thicknesses.

## IV. CONCLUSIONS

Small signal device simulations were performed to quantify the effect of temperature dependent substrate loss in GaN-on-Si transmission lines. The simulation methodology was validated using CPW lines on $SiN_x$ films on different Si substrate resistivities. The effect of the CPW dimensions and passivation layer thicknesses on substrate loss were also presented. The substrate loss contribution to total line loss in GaN-on-Si is not significant for temperatures up to 125°C but is shown to increase drastically for T >125°C in the GHz range on HR-Si substrates. On the contrary, low-resistivity Si substrates show a higher magnitude of loss and a crossover between an increasing and a decreasing trend with increasing temperatures as a function of frequency. While HR-Si delivers the lowest substrate loss for GaN-on-Si transmission lines across the considered temperature range for f>12GHz, very-low resistivity Si has a much lower and more temperature invariant substrate loss than even HR-Si for lower frequencies and is a promising substrate for wireless communication applications in terms of substrate loss considerations, with the provisos that line dimensions are re-designed for standard impedances, and appropriate matching circuits are used to compensate for the higher output capacitance (although this may reduce the achievable RF bandwidth).

## APPENDIX A

### THERMAL SIMULATIONS OF SUBSTRATE TEMPERATURE BELOW ACTIVE ALGAN/GAN TRANSISTORS

The impact of heat flow from the channel on substrate temperatures for an active transistor is illustrated here. Here we use a heatsink at 25°C ambient temperature, however at elevated heatsink temperatures, device self-heating will have a correspondingly bigger effect on substrate loss. Fig. 11 shows the results of a 3–D finite element thermal simulation of a representative AlGaN/GaN HEMT of 4.8 mm gate periphery (32×150 μm) and a generic 2 μm epitaxial stack on a 50 μm thinned HR-Si substrate, dissipating 4W/mm. This model was used to investigate temperature distribution within the Si of an operating device and consists of GaN (800 nm thickness, thermal conductivity, κ=160×300/T W/m·K), AlGaN strain relief layer (1 μm thick, κ=20×300/T W/m·K),[23] 25 μm-thick 35 W/m·K die attach and 1 mm-thick 200 W/m·K carrier. Surface metallization does not have a significant influence on the substrate temperature and was omitted. A 25°C isothermal boundary condition was applied at the back of the carrier; any additional temperature rise due to packaging thermal resistance should be added to the values here. We find that the peak channel temperature is 196°C and Si temperature reaches 153°C directly below the channel, decreasing to 105°C at the substrate back-side. As mentioned earlier, this is a lower limit temperature, excluding the thermal resistance of the packaging.

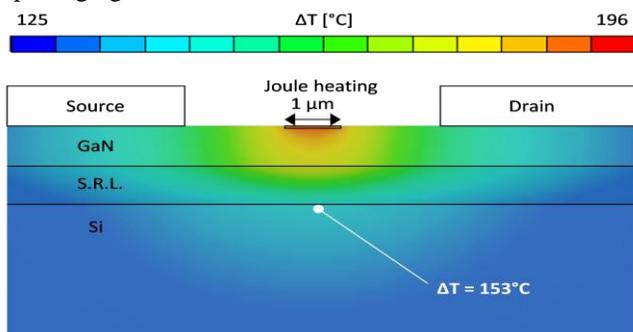

Fig. 9. Thermal simulations of a generic 4.8mm wide AlGaN/GaN HEMT stack on Si with a nominal power density of 4 W/mm, showing the temperature rise around the channel region and the extent of temperature rise in the substrate. The Joule heating region highlighted is based on Ref.[11].

## APPENDIX B

### TEMPERATURE DEPENDENT RESISTIVITY OF SILICON

To estimate the temperature dependence of silicon resistivity, the temperature dependence of the band gap, intrinsic carrier concentration and mobility were considered. The temperature dependence of the band-gap was calculated using the relation

$$E_G(T) = E_G(0) - \alpha T^2/(T+\beta), \quad (2)$$

with $E_G(0) = 1.169$ eV being the band gap at 0 K, and the constants $\alpha$ and $\beta$ having values of $4.9 \times 10^{-5}$ eV/K and 655 K respectively for Si. [24]

The intrinsic carrier concentration has a temperature dependence given by

$$n_i = \sqrt{(N_C N_V)} \exp(-E_G/2kT), \quad (3)$$

with $N_C$ and $N_V$ being the effective conduction and valence band density of states which in turn have a temperature dependence given by,

$$N_C = 2M_C(2\pi m_{de}kT/h^2)^{3/2}, \text{ and} \quad (4)$$

$$N_V = 2(2\pi m_{dh}kT/h^2)^{3/2}, \quad (5)$$

where $M_C$ is the number of equivalent conduction band minima (6 for Si), $m_{de}$ and $m_{dh}$ are the electron and hole density of states effective masses, k is the Boltzmann constant and h the Planck's constant.[24] The temperature dependence of mobility was estimated using the empirical relationship

$$\mu(T) = \mu(300K)*(300/T)^{2.2}, \quad (6)$$

where $\mu(300K)$ is the mobility at 300K for the different Si resistivities considered.[24]

Fig. 10 pictures the temperature-dependent resistivity of Si for three different room temperature resistivities (high-10kΩ, medium and low). We see that for HR-Si, the resistivity increases with temperature and peaks at 80 °C before dropping sharply which would translate to a mild decrease in substrate loss with temperature followed by a steep increase. In contrast, medium and low-resistivity material should exhibit an increase in resistivity (and decrease in substrate loss) with temperature up to 200 °C.

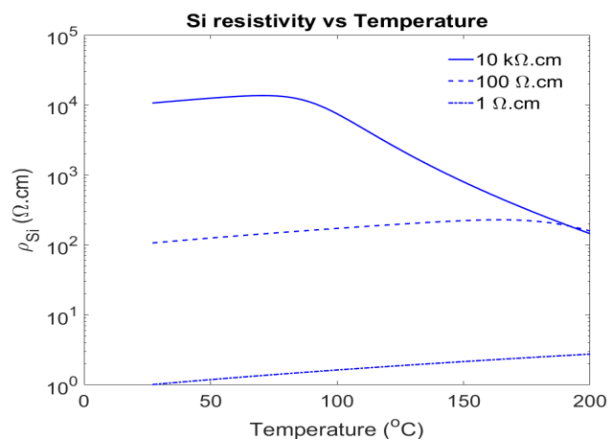

Fig. 10. Variation in silicon resistivity with temperature for high-resistivity (10 kΩ.cm), medium-resistivity (100 Ω.cm) and low-resistivity substrates (1

Ω.cm). Resistivity first increases and then drops sharply with temperature for high-res material while it increases for medium and low-res silicon.